\documentclass[aps,pra,twocolumn,showpacs]{revtex4-1}
\usepackage{graphicx,amsfonts}
\usepackage{amsmath,amscd,amssymb,color,bm}
\newcommand{\bra}[1]{\left\langle{#1}\right\vert}
\newcommand{\ket}[1]{\left\vert{#1}\right\rangle}
\begin{document}
\title{Experimental construction of a W-superposition
state and its equivalence to the GHZ state under local
filtration}
\author{Debmalya Das}
\email{debmalya@iisermohali.ac.in}
\affiliation{Department of Physical Sciences, Indian
Institute of Science Education \& 
Research Mohali, Sector 81 Mohali, 
Manauli PO 140306 Punjab India.}
\author{Shruti Dogra}
\email{shrutidogra@iisermohali.ac.in}
\affiliation{Department of Physical Sciences, Indian
Institute of Science Education \& 
Research Mohali, Sector 81 Mohali, 
Manauli PO 140306 Punjab India.}
\author{Kavita Dorai}
\email{kavita@iisermohali.ac.in}
\affiliation{Department of Physical Sciences, Indian
Institute of Science Education \& 
Research  Mohali, Sector 81 Mohali, 
Manauli PO 140306 Punjab India.}
\author{Arvind}
\email{arvind@iisermohali.ac.in}
\affiliation{Department of Physical Sciences, Indian
Institute of Science Education \& 
Research  Mohali, Sector 81 Mohali, 
Manauli PO 140306 Punjab India.}
\begin{abstract}
We experimentally construct a novel three-qubit entangled
W-superposition ($\rm W \bar{\rm W}$) state  on an NMR
quantum information processor.  We give a measurement-based
filtration protocol for the invertible local operation (ILO)
that converts the $\rm W \bar{\rm W}$ state to the GHZ
state, using a register of three ancilla qubits.  Further we
implement an experimental protocol to reconstruct full
information about the three-party $\rm W \bar{\rm W}$ state
using only two-party reduced density matrices.  An
intriguing fact unearthed recently is that the 
$\rm W \bar{\rm W}$ state which is 
equivalent to the GHZ state under ILO, is in fact
reconstructible from its two-party reduced density
matrices, unlike the GHZ state.  
We hence demonstrate that although the $\rm W \bar{\rm
W}$ state is interconvertible with the GHZ state, it stores
entanglement very differently.  
\end{abstract}
\maketitle
\section{Introduction} 
\label{intro} 
Explorations of multiqubit entanglement have unearthed
several families of states with curious quantum properties
and there have been many attempts in recent years to
characterize all the denizens of this
quantum zoo~\cite{horodecki-rmp-09,guhne-review,eltschka-jpa-14}.
The situation becomes complicated for systems of more than
two qubits and correspondingly the classification of their
entanglement turns out to be more
involved~\cite{vicente-prl-12,zhao-pra-13}.  

Pure entangled states of three qubits fall into two
categories, namely the GHZ- or the W-class, under stochastic
local operations and classical communication
(SLOCC)~\cite{acin-prl-00,kampermann-pra-12} with the
maximally entangled GHZ and W states being given by:
\begin{eqnarray}
\ket{\rm GHZ}&=&\frac{1}{\sqrt{2}}\left(\ket{000}
+\ket{111}\right) \nonumber \\
\vert {\rm W} \rangle &=& \frac{1}{\sqrt{3}}
\left( \vert 001 \rangle + \vert 010 \rangle
+ \vert 100\rangle \right)
\end{eqnarray}
The entanglement of the GHZ state is fragile under qubit
loss, i.e when any one of the qubits is traced out, the
other two qubits become completely
disentangled~\cite{chen-pra-06,guhne-review}.  Hence if one
of the parties decides not to cooperate, the entanglement
resources of the GHZ state cannot be used.  In
contradistinction to the GHZ state, the W-state residual
bipartite entanglement is robust against qubit
loss~\cite{guhne-review}.

It has been shown by Linden et.~al., that almost
every pure state of three qubits can be completely
determined by its two-party reduced density
matrices~\cite{linden-prl-1-02}.  The two
inequivalent entangled states, namely the W and
GHZ states, have contrasting irreducibility
features: while GHZ states have irreducible
correlations and cannot be determined from their
two-party
marginals~\cite{walck-prl-08,walck-pra-09},
W-states are completely determined by their
two-party marginals~\cite{diosi-pra-04,
cavalcanti-pra-05,parashar-pra-09}.  Tripartite
entanglement has been studied experimentally using
optics~\cite{roos-science-04,mikami-prl-05,
resch-prl-05} and
NMR~\cite{laflamme-proc-98,nelson-pra-00,
teklemariam-pra-02,kawamura-ijqc-06,
peng-pra-10,gao-pra-13, dogra-pra-15}.

Recently, the entanglement properties of a permutation
symmetric superposition of the $\rm W$ state and its obverse
$\bar{\rm W}=1/\sqrt{3}\left(\vert 100 \rangle + \vert 101
\rangle + \vert 110 \rangle \right)$ have been
characterized~\cite{ushadevi-qip-12,sudha-pra-12}:  
\begin{eqnarray}
&&\vert \rm W \bar{\rm W} \rangle =
\frac{1}{\sqrt{2}}\left(\ket{\rm W} + \ket{\bar{\rm
W}}\right)\nonumber \\
&&=\frac{1}{\sqrt{6}} \left(
\vert 001 \rangle 
+ \vert 010 \rangle + \vert 011 \rangle 
+ \vert 100 \rangle 
+ \vert 101 \rangle + \vert 110 \rangle \right) \nonumber \\
\end{eqnarray}
While this state (referred to henceforth as the $\rm
W\bar{\rm W}$ state) belongs to the GHZ entanglement class,
its correlation information (in contrast to the GHZ state)
is uniquely contained in its two-party reduced states.  The
argument for reconstructing the three-qubit $\rm W \bar{\rm
W}$ state from its two-party reduced states runs along
similar lines to the original argument of Linden
et.~al.~\cite{linden-prl-1-02}. If we assume another state
to have the same two-party reduced density matrices as the
$\rm W \bar{\rm W}$ state, this constraint can be used to
prove that the new state is no different from the original
$\rm W \bar{\rm W}$ state~\cite{ushadevi-qip-12,sudha-pra-12}.

In this work we focus on the $\rm W \bar{\rm W}$ state.  We
provide an explicit measurement-based filtration scheme to
filter out the $\ket{\rm GHZ}$ state from the $\rm W
\bar{\rm W}$ state.  Further, we experimentally construct
and tomograph the $\rm W \bar{\rm W}$ state on an NMR quantum
information processor of three coupled qubits.  We
experimentally demonstrate that the information about
tripartite correlations present in this state can indeed be
completely captured by its two-party reduced density
matrices.  We reconstruct the experimental density matrices
using complete state tomography and compare them with the
theoretically expected states and also compute state
fidelities.  The GHZ class of states are an important
computational resource~\cite{horodecki-rmp-09} and it has
been shown that states that are SLOCC equivalent to these can
be used for the same kind of quantum information processing
tasks~\cite{guhne-review}. Therefore, it is expected that the $\rm
W\bar{\rm W}$ state will also prove useful for quantum
computation. Furthermore, the quantification of the
tripartite correlation information present in this state is
easier as compared to the GHZ state, as entanglement
measurement requires only two-qubit detectors.

The paper is organized as follows: Section~\ref{filter}
describes how we obtain the ${\rm GHZ}$ state from the $\rm
W \bar{\rm W}$ state by local filtration based on projective
measurements using a register of three ancilla qubits.
Section~\ref{nmr} describes the experimental creation of the
$\rm W\bar{\rm W}$ superposition state on a three-qubit NMR
quantum information processor.  Section~\ref{nmr-wwbar}
contains the details of the molecule used, the NMR pulse
sequence for $\rm W \bar{\rm W}$ state construction and the
results of state tomography.  The information content of the
$\rm W\bar{\rm W}$  as captured from its two-party marginals
is described in Section~\ref{tomo-wwbar}.  We conclude in
Section~\ref{concl} with some remarks about GHZ
and W$\bar{\rm W}$
types of three-qubit entanglement and the relationship
between entanglement class and how information about
entanglement is stored in a quantum state.
\section{Filtration Protocol to show SLOCC 
equivalence of $\rm W\bar{\rm W}$ and GHZ} 
\label{filter}
Measurement-based local filters have been 
used for entanglement manipulation in the
context of violation of Bell inequalities
as well as for the detection of bound
entangled states~\cite{gisin-pla-96,prl-verstraete-02,das-qph-15}.
No local operations can convert a state from
the GHZ class to the W class.  However, surprisingly, it has
been shown that the $\rm{W} \bar{\rm W}$ is in the
GHZ class, deriving from the fact that it is related to
the GHZ state via the SLOCC class of operations given
by~\cite{ushadevi-qip-12,sudha-pra-12}:
\begin{equation}
\ket{GHZ} \equiv A\otimes A \otimes A \ket{W 
\bar{W}}
\label{relationship}
\end{equation}
with 
\begin{equation}
A=\frac{1}{\sqrt{3}}
\begin{pmatrix}
1 & \omega \\
1 & \omega^2
\end{pmatrix}
\label{rel}
\end{equation}
being an ILO, where
$\omega=e^{\iota \frac{2\pi}{3}}$ denotes the cube
root of unity.  We have used `$\equiv$' instead of
an equality sign in Eqn.~(\ref{relationship})
because $A$ is a non-unitary operator that does not
preserve the norm and the two sides in
Eqn.~(\ref{relationship}) do not have the same
norm.

We now proceed to reinterpret $A$ as an action on
an ensemble of identically prepared states $\rm
W\bar{\rm W}$ and implement the operation
described in  Eqn.~(\ref{relationship}). In
this process,  we will have to discard some copies
and the new ensemble that we construct with each
member in the filtered GHZ state will have fewer
copies as compared to the original ensemble of
$\rm W \bar{\rm W}$ states.  These aspects will be
brought out more clearly when we describe the
measurement-based filtration protocol to realize
the ILO.

Since $A$ acts on each of the qubits locally, we
first want to realize the operation $A$ on a
single qubit.  The non-unitary operator $A$ has a
singular valued decomposition 
\begin{equation}
 A=UDV
\label{singular-1}
\end{equation}
where the unitary operators $U$ and $V$ are given
by 
\begin{equation}
U = 
\frac{e^{\iota\frac{\pi}{2}}}{\sqrt{2}}
\begin{pmatrix}
e^{-\iota \frac{\pi}{6}} &    
-e^{\iota \frac{\pi}{3}}\\     
e^{\iota \frac{ \pi}{6}} &    
-e^{-\iota \frac{ \pi}{3}}     
       \end{pmatrix}, \quad
V =\frac{1}{\sqrt{2}}\begin{pmatrix}
       -\iota & \iota \\
       \iota & \iota
      \end{pmatrix} 
\end{equation}    
and the non-unitary diagonal operator $D$ is given by
\begin{equation}
D = \begin{pmatrix}
1 & 0\\
0 & \frac{1}{ \sqrt{3}}
\end{pmatrix}
\label{singular-2}
\end{equation}
The operators $U$ and $V$ are  unitary  and
can be implemented  via a local Hamiltonian
evolution. Therefore, we now turn to the
implementation of $D$ on a one-qubit state.

From the two columns of the operator $D$ we define
two vectors
\begin{equation}
\ket{u_1}=\begin{pmatrix}
1\\0
\end{pmatrix} 
\quad {\rm and} \quad  
\ket{u_2}=
\frac{1}{\sqrt{3}}
\begin{pmatrix}
0\\ 3^{\frac{1}{4}}
\end{pmatrix}
\end{equation}
These vectors are orthogonal to each other but are
not normalized. We now extend the Hilbert space of
the system by adding an ancilla qubit.  We extend
the vectors $u_1$ and $u_2$ to the composite
Hilbert space formed by the ancilla and the system
to obtain two four-dimensional vectors
\begin{equation}
\ket{\xi_1}=\begin{pmatrix}
1\\0\\0\\0
\end{pmatrix}
\quad
{\rm and} \quad 
\ket{\xi_2}= 
\frac{1}{\sqrt{3}}
\begin{pmatrix}
0\\ 3^{\frac{1}{4}}\\0\\ \sqrt{3-\sqrt{3}}
\end{pmatrix}
\end{equation}
The vectors $\ket{\xi_1}$ and $\ket{\xi_2}$ are
not only mutually orthogonal but also normalized.

Using these orthonormal vectors $\ket{\xi_1}$ and
$\ket{\xi_2}$, we construct
orthogonal projectors $P_1$ and $P_2$
\begin{eqnarray}
 P_1=  \ket{\xi_1}\bra{\xi_1} &=& \begin{pmatrix}
       1 & 0 & 0 & 0\\
       0 & 0 & 0 & 0\\
       0 & 0 & 0 & 0\\
       0 & 0 & 0 & 0
         \end{pmatrix}\nonumber\\
 P_2 = \ket{\xi_2}\bra{\xi_2}&=& 
\frac{1}{\sqrt{3}}
\begin{pmatrix}
        0 & 0 & 0 & 0\\
        0 & 1 & 0 & 
\sqrt{\sqrt{3}-1} 
\\
        0 & 0 & 0 & 0\\
        0 & 
\sqrt{\sqrt{3}-1} 
& 0 & \sqrt{3}-1
       \end{pmatrix}\nonumber \\
\label{projectors}
\end{eqnarray}
We define the projection operator $P=P_1+P_2$.
The effect of the projector $P$ on the composite
system of the single qubit and a one-qubit ancilla
turns out to be
\begin{equation}
P= \left(
\begin{array}{c|c}
D & \Delta \\\hline
\Delta & D^{\prime}
\end{array} \right)_{\displaystyle 4 \times 4}
\end{equation}
where $D$ is the diagonal part of the singular
value decomposition of the operator $A$ given in
Eqn.~(\ref{singular-2}), the complementary matrix
$D^{\prime}=I-D$ and the matrix $\Delta$ can be
obtained readily  from 
Eqn.~(\ref{projectors}).

If we prepare the ancilla in a state
$\ket{0}\bra{0}$ with the system being in an
arbitrary state $\rho$, the action of $P$ on the
composite system is given by
\begin{eqnarray}
 P\left(\ket{0}\bra{0}\otimes \rho\right) P 
= \left(
\begin{array}{c|c}
D\rho D & D\rho \Delta \\\hline
\Delta \rho D & \Delta \rho \Delta 
\end{array} \right)
\label{outcome}
\end{eqnarray}
\begin{figure}
\includegraphics[]{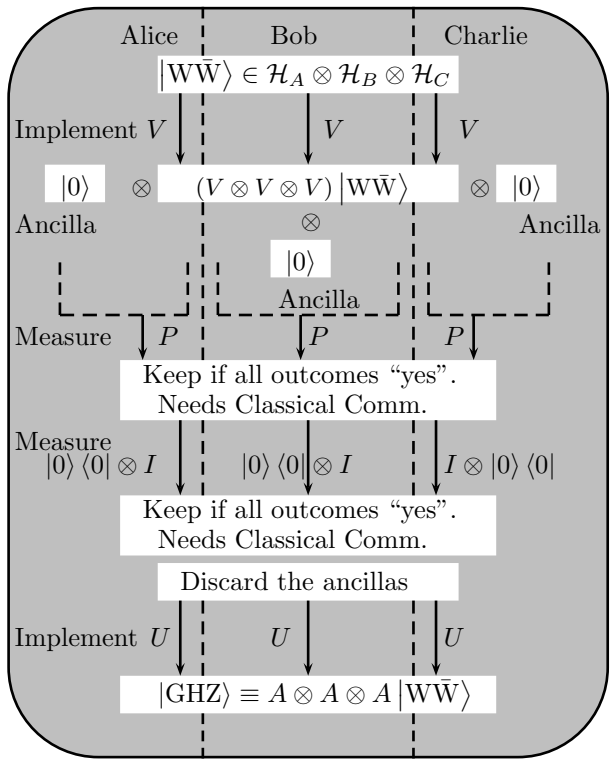}
\caption{Schematic diagram of the filtration
scheme to implement the non-unitary ILO
transformation that converts a $\rm W \bar{\rm W}$ 
state  to a  GHZ state.}
\label{schematic}
\end{figure}
If we measure the projector $P$ on the composite
system (system and ancilla), whenever the
measurement gives a positive answer, the state
after measurement is given by the right hand side
of Eqn.~(\ref{outcome}). We retain only these
cases and discard the state whenever the outcome
of the measurement is negative. Further, on the
final state given in Eqn.~(\ref{outcome}), we
measure the projector $\ket{0}\bra{0}$ on the
ancilla alone.  As before, if the outcome is
positive we retain the state, and if the outcome
is negative we discard the state.  In case the
outcome is positive, the resultant state is
$\ket{0}\bra{0} \otimes D \rho D$ and upon
discarding the ancilla we get the state of the
system to be $D \rho D$. This completes
the application of the non-unitary invertible
operator $D$ on $\rho$. Sandwiching this operation
between the unitary
transformations $U$ and $V$ as given in
Eqn.~(\ref{singular-1}), we achieve the
application of the ILO operator $A$ on $\rho$.

The scheme is easily extendable to $2\otimes 2
\otimes 2$ systems, where we locally implement $A$
on each of the three qubits. We imagine that the
tripartite system is divided between Alice, Bob
and Charlie and each of them can perform local
operations at their location. We begin with the
state $\ket{\rm W \bar{\rm W}}$ for the three
qubits,  attach a one-qubit ancilla to each qubit,
and measure the local projector $P$ for each
qubit.  If the outcome of these measurements (that
amount to a measurement of $P \otimes P \otimes
P$) is positive we retain the state, otherwise we
discard the state. Then on each ancilla, we
measure the projector $\ket{0} \bra{0}$ and retain
the cases when all the outcomes are positive. Upon
discarding the ancillas, the resultant state is
the application of $D$ on each qubit. When we
sandwich this process between
the unitaries $U$ and $V$ on each qubit, we get
the final state as $\ket{\rm GHZ}$.  This process
of measurement-based filtration is schematically
explained in Fig.~\ref{schematic}.  To decide when
to discard and when to retain the outcome, we
require classical communication between Alice, Bob
and Charlie. Since we discard
the output state in a number of cases, the size of
the ensemble obtained in the end is smaller than
the original ensemble.
\section{NMR implementation}
\label{nmr}
\begin{figure}[h]
\centering
\includegraphics[scale=1.0]{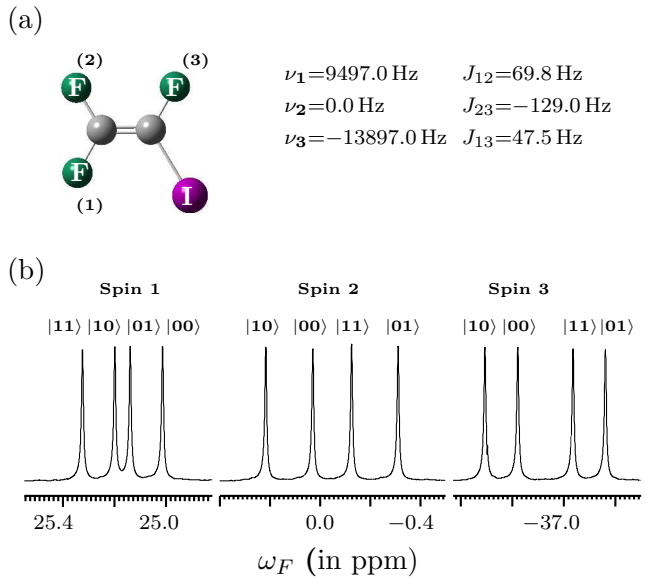}
\caption{
(a) Molecular structure and NMR parameters 
(chemical shifts and J-coupling in Hz)
and
${}^{19}$F NMR spectrum 
of trifluoroiodoethylene.
The three fluorine spins correspond to the
three-qubit system.
(b) The 
1D ${}^{19}$F NMR thermal equilibrium 
spectrum obtained after a $\frac{\pi}{2}$
readout pulse.
The NMR transitions of each qubit are labeled by
the corresponding logical states of the other
two qubits.
\label{molfig}
}
\end{figure}
To prepare the $\rm W \bar{\rm W}$ state on a three-qubit
NMR quantum information processor, we 
employ the three fluorine (spin-1/2) qubits of
trifluoroiodoethylene.
The molecular structure and
NMR parameters of this three-qubit system 
are adequate for the kind of manipulations involved
in quantum state preparation and are
given in Fig.~\ref{molfig}(a). 
Average fluorine longitudinal T$_1$ relaxation times of 
5.0 s and T$_2$ relaxation times of 1.0 s were
experimentally determined. The equilibrium  fluorine
NMR spectrum obtained after a $\frac{\pi}{2}$ readout pulse
is shown in Fig.~\ref{molfig}(b).

The system was first initialized into the $\vert 000
\rangle$ pseudopure state using the standard spatial
averaging technique~\cite{cory-physicad}.  The experimental
density matrices were tomographed by standard state
tomography
procedures~\cite{chuang-proc-98,long-joptb,leskowitz-pra-04}.
The three-qubit experimental density matrix was tomographed
using a set of eleven detection operators defined by \{III,
IIX, IXI, XII, IIY, IYI, YII, YYI, IXX, XXX, YYY\}, and the
two, two-qubit reduced density matrices were determined
using a set of four detection operators defined by \{III,
IXI, IYI, XXI\} and \{III, IIX, IIY, IXX\} respectively, with
I denoting the identity (or no-operation) operator and X(Y)
denoting a spin-selective $\frac{\pi}{2}$ pulse of X(Y)
phase on a specified qubit. 
The fidelity of the reconstructed state
was computed using the Uhlmann-Jozsa fidelity 
measure~\cite{uhlmann-fidelity,jozsa-fidelity}:
\begin{equation}
F =
\left(Tr \left( \sqrt{
\sqrt{\rho_{\rm theory}}
\rho_{\rm expt} \sqrt{\rho_{\rm theory}}
}
\right)\right)^2
\label{fidelity}
\end{equation}
where $\rho_{\rm theory}$ and $\rho_{\rm expt}$ denote the
theoretical and experimental density matrices
respectively.
\subsection{$\rm W\bar{\rm W}$ construction scheme}
\label{nmr-wwbar}
The circuit to construct a $\rm W \bar{\rm W}$ state
consists of several single-qubit and two-qubit gates.
A single-qubit gate $\rm U_{i}[\alpha]_y$ acting
on the $i$th qubit, achieves a rotation by the
angle $\alpha$ around the $y$ axis with a corresponding
unitary matrix given by:
\begin{equation}
U_{i}[\alpha]_y = \left(\begin{array}{cc}
\cos{\frac{\alpha}{2}} & - \sin{\frac{\alpha}{2}} \\
\sin{\frac{\alpha}{2}} & \cos{\frac{\alpha}{2}}
\end{array}
\right)
\end{equation}
A two-qubit controlled-rotation gate CR$_{ij}[\phi]_y$,
implements the single-qubit rotation $U_j[\phi]_y$ on
the target qubit $j$ about the $y$ axis,
if the  control qubit $i$ is in the
state $\vert 1 \rangle$.
The $\rm CNOT_{ij}$ gate implements a controlled-NOT
operation with the $i$th qubit as control and the
$j$th qubit as target. 

\begin{figure}[h]
\includegraphics[scale=1.0]{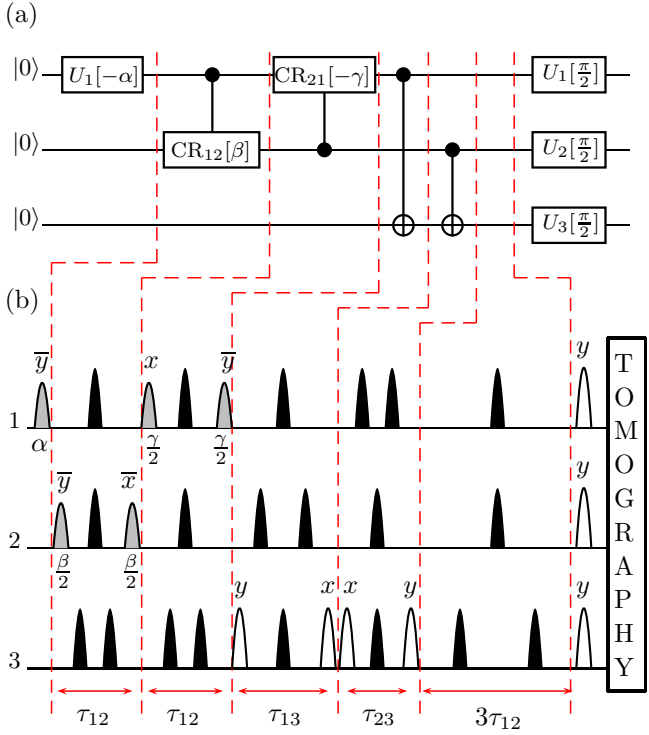}
\caption{(a) Quantum circuit showing
sequence of gates required to construct 
the $\rm W \bar{\rm W}$ state, starting
from the pseudopure state $\vert 000 \rangle$.
The gate operations are described in the
main text and all the rotations take place about the $y$-axis.
(b) NMR pulse sequence to create a 
$\rm W \bar{\rm W}$ state.
All the pulses are low-power selective pulses
represented by shaped blocks. Filled black shapes are 
$\pi$ refocusing pulses, unfilled shapes correspond to
pulses of $\frac{\pi}{2}$ flip angle and the gray
shaded shapes are 
labeled with their specific flip angles and phases.
The axes of
rotation are specified at the top of each pulse. Vertical
dotted red lines show the correspondence between the
quantum circuit and the experimental pulse sequence. 
All pulses are of 
phase $x$ unless otherwise labeled. 
The values of the rf pulse flip angles used are
$\alpha=\frac{\pi}{3}, \beta=2 \cos^{-1}{(\frac{1}{\sqrt{3}})}, 
\gamma=\frac{\pi}{2}$ and
$\tau_{ij}$ represents an evolution under
the $J_{ij}$ coupling. The last $3 \tau_{12}$
period
is used to compensate the extra phase
acquired (as described in the text). 
\label{pulsefig}
}
\end{figure}

The sequence of gates to construct a $\rm W\bar{\rm W}$
state, starting from the initial pseudopure
state $\vert 0 0 0 \rangle$ is given as:
\begin{eqnarray}
&\vert 0 0 0 \rangle &\nonumber \\ &
\downarrow
\framebox[1.25cm]
{
$\scriptstyle U_{1}\left[-\frac{\pi}{3}\right]_y$
} 
\downarrow 
&\nonumber \\&
\frac{1}{2}\left(
\sqrt{3} \vert 0 0 0 \rangle -
\vert 1 0 0 \rangle \right)
&\nonumber \\&
\downarrow
\framebox[3.0cm]
{
$\scriptstyle {\rm
CR}_{12}\left[2\cos^{-1}{(1/\sqrt{3})}\right]_y$} 
\downarrow
&\nonumber \\&
\frac{1}{2}\left(
\sqrt{3}\vert 0 0 0 \rangle -
\frac{1}{\sqrt{3}} \vert 1 0 0 \rangle -
\sqrt{\frac{2}{3}} \vert 1 1 0 \rangle 
\right)
&\nonumber \\&
\downarrow{\framebox[1.5cm]
{$ \scriptstyle {\rm
CR}_{21}\left[-\frac{\pi}{2}\right]_y$}}
\downarrow
&\nonumber \\&
\frac{1}{2}\left(
\sqrt{3}\vert 0 0 0 \rangle -
\frac{1}{\sqrt{3}} \left( \vert 1 0 0 \rangle  
+\vert 1 1 0 \rangle+\vert 0 1 0 \rangle 
\right)\right)
&\nonumber \\&
\downarrow{\framebox[1.5cm]
{$\scriptstyle{\rm CNOT}_{13}$}} 
\downarrow
&\nonumber \\ &
\frac{1}{2}\left(
\sqrt{3}\vert 0 0 0 \rangle -
\frac{1}{\sqrt{3}} \left(\vert 1 0 1 \rangle 
+ \vert 1 1 1 \rangle + \vert 0 1 0 \rangle 
\right)\right)
&\nonumber \\ &
\downarrow{\framebox[1.5cm]
{$\scriptstyle {\rm CNOT}_{23}$}} 
\downarrow
& \nonumber \\ &
\frac{1}{2}\left(
\sqrt{3} \vert 0 0 0 \rangle -
\frac{1}{\sqrt{3}} \left( \vert 1 0 1 \rangle 
 + \vert 1 1 0 \rangle +
\vert 0 1 1 \rangle 
\right)\right)
& \nonumber \\ &
\downarrow{\framebox[3cm]
{$\scriptstyle 
\rm U_1\left[\frac{\pi}{2}\right]_y
\rm U_2\left[\frac{\pi}{2}\right]_y
\rm U_3\left[\frac{\pi}{2}\right]_y
$}} 
\downarrow
& \nonumber \\ &
\frac{1}{\sqrt{6}} \left(\vert 0 0 1 \rangle + \vert 0 1 0 \rangle
+ \vert 0 1 1 \rangle   
+ \vert 1 0 0 \rangle + 
\vert 1 0 1 \rangle + \vert 1 1 0 \rangle \right) 
& \nonumber 
\end{eqnarray}
The quantum
circuit to construct the $\rm W \bar{\rm W}$ state
on a three-qubit system is given in Fig.~\ref{pulsefig}(a).
The NMR pulse sequence to create the $\rm W \bar{\rm
W}$ state, starting from the pseudopure state $\vert 000
\rangle$ is given in Fig.~\ref{pulsefig}(b). All the
pulses are shaped pulses, labeled by the corresponding axes
of rotation and the flip angles; $\tau_{ij}$ denotes an
evolution period under the $J_{ij}$ coupling. Refocusing
($\pi$) pulses are applied in the middle of the evolution
periods to compensate for chemical shift evolution and
pairs of $\pi$ pulses are introduced at $1/4$ and $3/4$ of
the evolution periods to eliminate undesired J-evolutions.
After the evolution interval $\tau_{23}$ and the
$[\frac{\pi}{2}]_y$ on
the third qubit (corresponding to a $\rm CNOT_{23}$ gate), the
state obtained is $\frac{\sqrt{3}}{2} \vert 0 0 0 \rangle -
\frac{1}{2\sqrt{3}} (\iota \vert 1 0 1 \rangle + \vert 1 1 0
\rangle + \iota \vert 0 1 1 \rangle )$.  There is an
undesirable extra relative phase of `$\iota$' that has
accumulated between two of the basis vectors.  This
undesired extra phase factor is compensated for during the
evolution interval $3\tau_{12}$.  The implementation of the
last module (simultaneous $[\frac{\pi}{2}]_y$ pulses on all the three
qubits) results in the desired $\rm W \bar{\rm W}$
state with no extra relative phase.  All the selective
pulses are $265~\mu$s ``{\em Gauss}'' shaped pulses and the
non-selective excitation pulse is a frequency-modulated
$400~\mu$s ``{\em Gauss}'' shaped pulse.
\begin{figure}[h]
\includegraphics[scale=1]{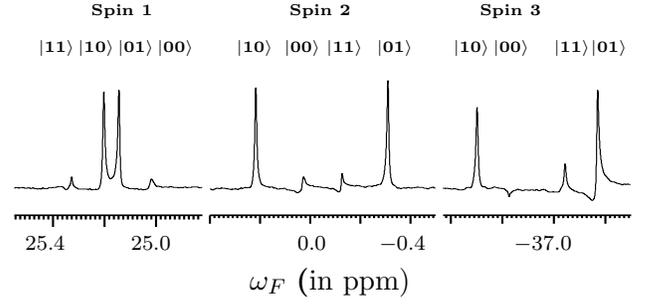}
\caption{ 
The 1D ${}^{19}$F NMR spectrum  corresponding to the
creation of the
$\rm W \bar{\rm W}$ state.
The NMR transitions of each qubit are labeled by
the corresponding logical states of the other
two qubits.
\label{spectrum}}
\end{figure}
\begin{figure}[h]
\includegraphics[scale=1]{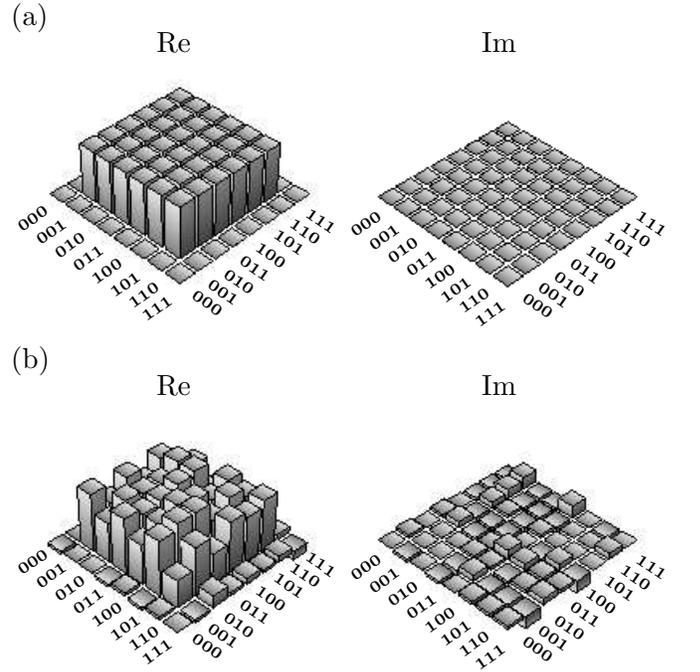}
\caption{ 
The real (Re) and imaginary (Im) parts of the
(a) theoretically expected and (b) experimentally
density matrices for the
$\rm W \bar{\rm W}$ state reconstructed
using full state tomography. The rows and columns of
the bar graphs depict the computational basis of the
three qubits in binary order from
$\vert 000\rangle$ to $\vert 111 \rangle$. The
experimentally tomographed state has a 
fidelity of 0.94.
\label{3tomo}
}
\end{figure}
The NMR spectrum of the $\rm W \bar{\rm W}$ state obtained
by a sequence of selective rotations on the initial
pseudopure state is shown in Fig.~\ref{spectrum}.  Each spin
multiplet has two resonance peaks (as compared to four
resonance peaks for the thermal equilibrium state).  The
expected NMR spectral pattern of an ideal $\rm W \bar{\rm
W}$ state should contain resonance peaks of equal magnitude
and phase, and deviations from ideal spectral peak
intensities and phases in the experimentally obtained
spectrum, can be attributed to imperfections in the rf pulse
calibrations and to relaxation during the selective pulse
durations.

The tomograph of the experimentally constructed $\rm W
\bar{\rm W}$ state is shown in Fig.~\ref{3tomo}. The
experimentally tomographed state was compared with the
theoretically expected state and the density matrices match
well, within experimental error, with a computed state
fidelity of 0.94 (the fidelity was computed from
Eqn.~\ref{fidelity}).
\subsection{Reconstruction of $\rm W\bar{\rm W}$ from
two-party reduced density matrices}
\label{tomo-wwbar}
A protocol was developed~\cite{diosi-pra-04}  to validate
the surprising aspect of multi-party correlations asserted
by Linden et.~al.~\cite{linden-prl-1-02,linden-prl-2-02},
that the information about three-party correlations  of
almost all pure three-qubit states (except for GHZ-type
states) are already contained in their corresponding
two-party reduced states.  We delineate below the
argument for how a general three-qubit pure state
$\rho_{ABC}$ can be completely determined by using
any of the equivalent sets $(\rho_{AB},\rho_{AC})$,
$(\rho_{AB},\rho_{BC})$, or $(\rho_{AC},\rho_{BC})$ of
reduced two-party states.  
The reduced single-qubit reduced state $\rho_{A}$ and 
the two-qubit reduced state $\rho_{BC}$ share the same
set of eigen values, and can hence be 
written as~\cite{diosi-pra-04}:
\begin{eqnarray}
\rho_A &=&\sum_i
p_A^i \ket{i} \bra{i}  \nonumber \\ 
\rho_{BC} &=& \sum_i p_A^i \ket{i;BC} \bra{i;BC}
\end{eqnarray}
where $\{\ket{i}\}$ are the eigenvectors of
$\rho_{A}$ with eigenvalues $\{p_A^i\}$,
and $\{ \ket{i;BC}\}$ are the eigenvectors of
$\rho_{BC}$ with eigenvalues $\{ p_A^i\}$.
Furthermore, the three-qubit pure states that are compatible
with $\rho_{A}$ and $\rho_{BC}$ are given by:
\begin{equation}
\ket{\psi_{ABC};\alpha}=\sum_i
e^{\iota \alpha_i}
\sqrt{p_A^i}\ket{i}\otimes\ket{i;BC}
\end{equation}
Similarly, the three-qubit pure states that are compatible
with $\rho_C$  and $\rho_{AB}$ are given by
\begin{equation}
\ket{\psi_{ABC};\gamma}=\sum_k
e^{\iota \gamma_k}
\sqrt{p_c^k}\ket{k;AB}\otimes\ket{k} 
\end{equation}
where $\{\ket{k}\}$ are the eigenvectors of $\rho_C$ with
eigenvalues $\{p_c^k\}$ and $\{\ket{k;AB}\}$ are the
corresponding eigenvectors of $\rho_{AB}$.  Since the pure
state $\ket{\psi_{ABC}}$ is compatible with both $\rho_{AB}$
and $\rho_{BC}$, we can now consistently find the values of
$\alpha_i$ and $\gamma_k$ while ensuring that
$\ket{\psi_{ABC};\alpha}=\ket{\psi_{ABC};\gamma}$.

\begin{figure}[h]
\includegraphics[scale=1]{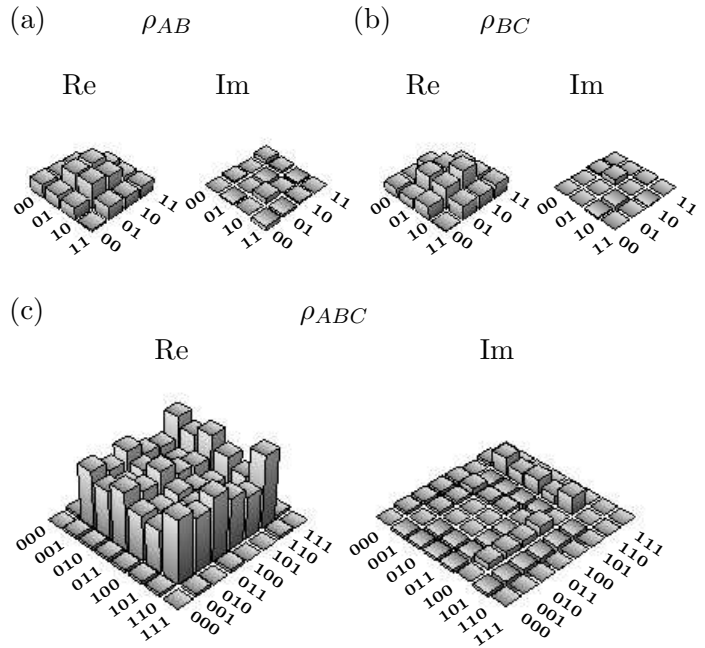}
\caption{
The real (Re) and imaginary (Im) parts of the
tomographed density matrix for the
state $\rm W \bar{\rm W}$ state. (a) The two-qubit
reduced density matrix $\rho_{AB}$.
(b) The two-qubit reduced density matrix
$\rho_{BC}$. (c) The entire three-qubit
density matrix $\rho_{ABC}$, reconstructed
from the corresponding two-qubit reduced density matrices.
The rows and columns in the bar graphs encode
the computational basis of the qubits, from
$\vert 00 \rangle$ to $\vert 11 \rangle$ for
two qubits and from $\vert 000 \rangle$ to
$\vert 111 \rangle$ for three qubits. 
The fidelity
between the three-qubit state ($\rho_{ABC}$) reconstructed
from the two-qubit density matrices
and the three-qubit state obtained by complete
three-qubit state tomography is found to
be 0.92.
\label{2tomo}
}
\end{figure}

We used the set of two, two-party reduced
states $(\rho_{AB},\rho_{BC})$, to
reconstruct the full three-qubit
$\rm W \bar{\rm W}$ state.
The reconstructed density matrix for the $\rm W
\bar{\rm W}$ state, using two sets of the corresponding
two-qubit reduced density matrices ($\rho_{AB},\rho_{BC}$)
is given in Fig.~\ref{2tomo}.  The 
two-party reduced states were able to reconstruct the
three-party $\rm W \bar{\rm W}$ state with a
fidelity of 0.92, which matches well with the
full reconstruction of the entire three-qubit state
given in Fig.~\ref{3tomo}(b).
\section{Conclusions}
\label{concl}
We described a measurement-based filtration scheme to
demonstrate the ILO equivalence of the $\rm W \bar{\rm W}$
state with the $\rm GHZ$ state.  We experimentally implemented an
NMR-based scheme to construct a $\rm W \bar{\rm W}$ state.
We were able to show that the three-qubit density operator
$\rho_{ABC}$ obtained by full state tomography matches well
with the same three-qubit state reconstructed using a set of
two-party reduced density operators $(\rho_{AB},
\rho_{BC})$.  Thus, although the $\rm W \bar{\rm W}$ state
belongs to the same entanglement class as the GHZ state, the
two states store information about multi-party correlations
in completely different ways.  We thus experimentally
demonstrated an interesting feature of multi-qubit
entanglement namely, that two different entangled states
belonging to the same SLOCC class can yet have their
correlations exhibiting contrasting irreducible properties.

Since distinguishing entangled states is still a hard task,
our work can be used as a benchmark to further classify how
different entangled states store information about their
correlations.  Our work also has important implications for
comparing the utility of different kinds of entangled states
to perform the same computational task.  We were unable to
find a suitable molecular architecture to experimentally
implement the ILO, since this requires each of the three
qubits to be coupled to a separate one-qubit ancilla.
However, it is a worthwhile exercise to look for an
experimental implementation of the filtering protocol to
perform the ILO. A further issue with such an implementation
is the involvement of projective measurements, which are not
straightforward to achieve using NMR. 
\begin{acknowledgments}
All experiments were performed on a Bruker Avance-III
400 MHz FT-NMR spectrometer at the NMR Research Facility
at IISER Mohali. SD acknowledges financial
support from UGC India. 
\end{acknowledgments}

\begin{thebibliography}{36}%
\makeatletter
\providecommand \@ifxundefined [1]{%
 \@ifx{#1\undefined}
}%
\providecommand \@ifnum [1]{%
 \ifnum #1\expandafter \@firstoftwo
 \else \expandafter \@secondoftwo
 \fi
}%
\providecommand \@ifx [1]{%
 \ifx #1\expandafter \@firstoftwo
 \else \expandafter \@secondoftwo
 \fi
}%
\providecommand \natexlab [1]{#1}%
\providecommand \enquote  [1]{``#1''}%
\providecommand \bibnamefont  [1]{#1}%
\providecommand \bibfnamefont [1]{#1}%
\providecommand \citenamefont [1]{#1}%
\providecommand \href@noop [0]{\@secondoftwo}%
\providecommand \href [0]{\begingroup \@sanitize@url \@href}%
\providecommand \@href[1]{\@@startlink{#1}\@@href}%
\providecommand \@@href[1]{\endgroup#1\@@endlink}%
\providecommand \@sanitize@url [0]{\catcode `\\12\catcode `\$12\catcode
  `\&12\catcode `\#12\catcode `\^12\catcode `\_12\catcode `\%12\relax}%
\providecommand \@@startlink[1]{}%
\providecommand \@@endlink[0]{}%
\providecommand \url  [0]{\begingroup\@sanitize@url \@url }%
\providecommand \@url [1]{\endgroup\@href {#1}{\urlprefix }}%
\providecommand \urlprefix  [0]{URL }%
\providecommand \Eprint [0]{\href }%
\providecommand \doibase [0]{http://dx.doi.org/}%
\providecommand \selectlanguage [0]{\@gobble}%
\providecommand \bibinfo  [0]{\@secondoftwo}%
\providecommand \bibfield  [0]{\@secondoftwo}%
\providecommand \translation [1]{[#1]}%
\providecommand \BibitemOpen [0]{}%
\providecommand \bibitemStop [0]{}%
\providecommand \bibitemNoStop [0]{.\EOS\space}%
\providecommand \EOS [0]{\spacefactor3000\relax}%
\providecommand \BibitemShut  [1]{\csname bibitem#1\endcsname}%
\let\auto@bib@innerbib\@empty
\bibitem [{\citenamefont {Horodecki}\ \emph {et~al.}(2009)\citenamefont
  {Horodecki}, \citenamefont {Horodecki}, \citenamefont {Horodecki},\ and\
  \citenamefont {Horodecki}}]{horodecki-rmp-09}%
  \BibitemOpen
  \bibfield  {author} {\bibinfo {author} {\bibfnamefont {R.}~\bibnamefont
  {Horodecki}}, \bibinfo {author} {\bibfnamefont {P.}~\bibnamefont
  {Horodecki}}, \bibinfo {author} {\bibfnamefont {M.}~\bibnamefont
  {Horodecki}}, \ and\ \bibinfo {author} {\bibfnamefont {K.}~\bibnamefont
  {Horodecki}},\ }\href@noop {} {\bibfield  {journal} {\bibinfo  {journal}
  {Rev. Mod. Phys.}\ }\textbf {\bibinfo {volume} {81}},\ \bibinfo {pages} {865}
  (\bibinfo {year} {2009})}\BibitemShut {NoStop}%
\bibitem [{\citenamefont {G\"uhne}\ and\ \citenamefont
  {T\"oth}(2009)}]{guhne-review}%
  \BibitemOpen
  \bibfield  {author} {\bibinfo {author} {\bibfnamefont {O.}~\bibnamefont
  {G\"uhne}}\ and\ \bibinfo {author} {\bibfnamefont {G.}~\bibnamefont
  {T\"oth}},\ }\href@noop {} {\bibfield  {journal} {\bibinfo  {journal}
  {Physics Reports}\ }\textbf {\bibinfo {volume} {474}},\ \bibinfo {pages} {1}
  (\bibinfo {year} {2009})}\BibitemShut {NoStop}%
\bibitem [{\citenamefont {Eltschka}\ and\ \citenamefont
  {Siewert}(2014)}]{eltschka-jpa-14}%
  \BibitemOpen
  \bibfield  {author} {\bibinfo {author} {\bibfnamefont {C.}~\bibnamefont
  {Eltschka}}\ and\ \bibinfo {author} {\bibfnamefont {J.}~\bibnamefont
  {Siewert}},\ }\href@noop {} {\bibfield  {journal} {\bibinfo  {journal} {J.
  Phys. A}\ }\textbf {\bibinfo {volume} {47}},\ \bibinfo {pages} {424005}
  (\bibinfo {year} {2014})}\BibitemShut {NoStop}%
\bibitem [{\citenamefont {de~Vicente}\ \emph {et~al.}(2012)\citenamefont
  {de~Vicente}, \citenamefont {Carle}, \citenamefont {Streitberger},\ and\
  \citenamefont {Kraus}}]{vicente-prl-12}%
  \BibitemOpen
  \bibfield  {author} {\bibinfo {author} {\bibfnamefont {J.~I.}\ \bibnamefont
  {de~Vicente}}, \bibinfo {author} {\bibfnamefont {T.}~\bibnamefont {Carle}},
  \bibinfo {author} {\bibfnamefont {C.}~\bibnamefont {Streitberger}}, \ and\
  \bibinfo {author} {\bibfnamefont {B.}~\bibnamefont {Kraus}},\ }\href@noop {}
  {\bibfield  {journal} {\bibinfo  {journal} {Phys. Rev. Lett.}\ }\textbf
  {\bibinfo {volume} {108}},\ \bibinfo {pages} {060501} (\bibinfo {year}
  {2012})}\BibitemShut {NoStop}%
\bibitem [{\citenamefont {Zhao}\ \emph {et~al.}(2013)\citenamefont {Zhao},
  \citenamefont {Zhang}, \citenamefont {Li-Jost},\ and\ \citenamefont
  {Fei}}]{zhao-pra-13}%
  \BibitemOpen
  \bibfield  {author} {\bibinfo {author} {\bibfnamefont {M.-J.}\ \bibnamefont
  {Zhao}}, \bibinfo {author} {\bibfnamefont {T.-G.}\ \bibnamefont {Zhang}},
  \bibinfo {author} {\bibfnamefont {X.}~\bibnamefont {Li-Jost}}, \ and\
  \bibinfo {author} {\bibfnamefont {S.-M.}\ \bibnamefont {Fei}},\ }\href@noop
  {} {\bibfield  {journal} {\bibinfo  {journal} {Phys. Rev. A}\ }\textbf
  {\bibinfo {volume} {87}},\ \bibinfo {pages} {012316} (\bibinfo {year}
  {2013})}\BibitemShut {NoStop}%
\bibitem [{\citenamefont {Acin}\ \emph {et~al.}(2000)\citenamefont {Acin},
  \citenamefont {Andrianov}, \citenamefont {Costa}, \citenamefont {Jane},
  \citenamefont {Latorre},\ and\ \citenamefont {Tarrach}}]{acin-prl-00}%
  \BibitemOpen
  \bibfield  {author} {\bibinfo {author} {\bibfnamefont {A.}~\bibnamefont
  {Acin}}, \bibinfo {author} {\bibfnamefont {A.}~\bibnamefont {Andrianov}},
  \bibinfo {author} {\bibfnamefont {L.}~\bibnamefont {Costa}}, \bibinfo
  {author} {\bibfnamefont {E.}~\bibnamefont {Jane}}, \bibinfo {author}
  {\bibfnamefont {J.~I.}\ \bibnamefont {Latorre}}, \ and\ \bibinfo {author}
  {\bibfnamefont {R.}~\bibnamefont {Tarrach}},\ }\href@noop {} {\bibfield
  {journal} {\bibinfo  {journal} {Phys. Rev. Lett.}\ }\textbf {\bibinfo
  {volume} {85}},\ \bibinfo {pages} {1560} (\bibinfo {year}
  {2000})}\BibitemShut {NoStop}%
\bibitem [{\citenamefont {Kampermann}\ \emph {et~al.}(2012)\citenamefont
  {Kampermann}, \citenamefont {G\"uhne}, \citenamefont {Wilmott},\ and\
  \citenamefont {Bru\ss{}}}]{kampermann-pra-12}%
  \BibitemOpen
  \bibfield  {author} {\bibinfo {author} {\bibfnamefont {H.}~\bibnamefont
  {Kampermann}}, \bibinfo {author} {\bibfnamefont {O.}~\bibnamefont {G\"uhne}},
  \bibinfo {author} {\bibfnamefont {C.}~\bibnamefont {Wilmott}}, \ and\
  \bibinfo {author} {\bibfnamefont {D.}~\bibnamefont {Bru\ss{}}},\ }\href@noop
  {} {\bibfield  {journal} {\bibinfo  {journal} {Phys. Rev. A}\ }\textbf
  {\bibinfo {volume} {86}},\ \bibinfo {pages} {032307} (\bibinfo {year}
  {2012})}\BibitemShut {NoStop}%
\bibitem [{\citenamefont {Chen}\ and\ \citenamefont
  {Chen}(2006)}]{chen-pra-06}%
  \BibitemOpen
  \bibfield  {author} {\bibinfo {author} {\bibfnamefont {L.}~\bibnamefont
  {Chen}}\ and\ \bibinfo {author} {\bibfnamefont {Y.~X.}\ \bibnamefont
  {Chen}},\ }\href@noop {} {\bibfield  {journal} {\bibinfo  {journal} {Phys.
  Rev. A}\ }\textbf {\bibinfo {volume} {74}},\ \bibinfo {pages} {062310}
  (\bibinfo {year} {2006})}\BibitemShut {NoStop}%
\bibitem [{\citenamefont {Linden}\ \emph {et~al.}(2002)\citenamefont {Linden},
  \citenamefont {Popescu},\ and\ \citenamefont {Wootters}}]{linden-prl-1-02}%
  \BibitemOpen
  \bibfield  {author} {\bibinfo {author} {\bibfnamefont {N.}~\bibnamefont
  {Linden}}, \bibinfo {author} {\bibfnamefont {S.}~\bibnamefont {Popescu}}, \
  and\ \bibinfo {author} {\bibfnamefont {W.~K.}\ \bibnamefont {Wootters}},\
  }\href@noop {} {\bibfield  {journal} {\bibinfo  {journal} {Phys. Rev. Lett.}\
  }\textbf {\bibinfo {volume} {89}},\ \bibinfo {pages} {207901} (\bibinfo
  {year} {2002})}\BibitemShut {NoStop}%
\bibitem [{\citenamefont {Walck}\ and\ \citenamefont
  {Lyons}(2008)}]{walck-prl-08}%
  \BibitemOpen
  \bibfield  {author} {\bibinfo {author} {\bibfnamefont {S.~N.}\ \bibnamefont
  {Walck}}\ and\ \bibinfo {author} {\bibfnamefont {D.~W.}\ \bibnamefont
  {Lyons}},\ }\href@noop {} {\bibfield  {journal} {\bibinfo  {journal} {Phys.
  Rev. Lett.}\ }\textbf {\bibinfo {volume} {100}},\ \bibinfo {pages} {050501}
  (\bibinfo {year} {2008})}\BibitemShut {NoStop}%
\bibitem [{\citenamefont {Walck}\ and\ \citenamefont
  {Lyons}(2009)}]{walck-pra-09}%
  \BibitemOpen
  \bibfield  {author} {\bibinfo {author} {\bibfnamefont {S.~N.}\ \bibnamefont
  {Walck}}\ and\ \bibinfo {author} {\bibfnamefont {D.~W.}\ \bibnamefont
  {Lyons}},\ }\href@noop {} {\bibfield  {journal} {\bibinfo  {journal} {Phys.
  Rev. A}\ }\textbf {\bibinfo {volume} {79}},\ \bibinfo {pages} {032326}
  (\bibinfo {year} {2009})}\BibitemShut {NoStop}%
\bibitem [{\citenamefont {Diosi}(2004)}]{diosi-pra-04}%
  \BibitemOpen
  \bibfield  {author} {\bibinfo {author} {\bibfnamefont {L.}~\bibnamefont
  {Diosi}},\ }\href@noop {} {\bibfield  {journal} {\bibinfo  {journal} {Phys.
  Rev. A}\ }\textbf {\bibinfo {volume} {70}},\ \bibinfo {pages} {010302}
  (\bibinfo {year} {2004})}\BibitemShut {NoStop}%
\bibitem [{\citenamefont {Cavalcanti}\ \emph {et~al.}(2005)\citenamefont
  {Cavalcanti}, \citenamefont {Cioletti},\ and\ \citenamefont
  {Cunha}}]{cavalcanti-pra-05}%
  \BibitemOpen
  \bibfield  {author} {\bibinfo {author} {\bibfnamefont {D.}~\bibnamefont
  {Cavalcanti}}, \bibinfo {author} {\bibfnamefont {L.~M.}\ \bibnamefont
  {Cioletti}}, \ and\ \bibinfo {author} {\bibfnamefont {M.~O.~T.}\ \bibnamefont
  {Cunha}},\ }\href@noop {} {\bibfield  {journal} {\bibinfo  {journal} {Phys.
  Rev. A}\ }\textbf {\bibinfo {volume} {71}},\ \bibinfo {pages} {014301}
  (\bibinfo {year} {2005})}\BibitemShut {NoStop}%
\bibitem [{\citenamefont {Parashar}\ and\ \citenamefont
  {Rana}(2009)}]{parashar-pra-09}%
  \BibitemOpen
  \bibfield  {author} {\bibinfo {author} {\bibfnamefont {P.}~\bibnamefont
  {Parashar}}\ and\ \bibinfo {author} {\bibfnamefont {S.}~\bibnamefont
  {Rana}},\ }\href@noop {} {\bibfield  {journal} {\bibinfo  {journal} {Phys.
  Rev. A}\ }\textbf {\bibinfo {volume} {80}},\ \bibinfo {pages} {012319}
  (\bibinfo {year} {2009})}\BibitemShut {NoStop}%
\bibitem [{\citenamefont {Roos}\ \emph {et~al.}(2004)\citenamefont {Roos},
  \citenamefont {Riebe}, \citenamefont {Haffner}, \citenamefont {Hansel},
  \citenamefont {Benhelm}, \citenamefont {Lancaster}, \citenamefont {Becher},
  \citenamefont {Schmidt-Kaler},\ and\ \citenamefont
  {Blatt}}]{roos-science-04}%
  \BibitemOpen
  \bibfield  {author} {\bibinfo {author} {\bibfnamefont {C.~F.}\ \bibnamefont
  {Roos}}, \bibinfo {author} {\bibfnamefont {M.}~\bibnamefont {Riebe}},
  \bibinfo {author} {\bibfnamefont {H.}~\bibnamefont {Haffner}}, \bibinfo
  {author} {\bibfnamefont {W.}~\bibnamefont {Hansel}}, \bibinfo {author}
  {\bibfnamefont {J.}~\bibnamefont {Benhelm}}, \bibinfo {author} {\bibfnamefont
  {G.~P.~T.}\ \bibnamefont {Lancaster}}, \bibinfo {author} {\bibfnamefont
  {C.}~\bibnamefont {Becher}}, \bibinfo {author} {\bibfnamefont
  {F.}~\bibnamefont {Schmidt-Kaler}}, \ and\ \bibinfo {author} {\bibfnamefont
  {R.}~\bibnamefont {Blatt}},\ }\href@noop {} {\bibfield  {journal} {\bibinfo
  {journal} {Science}\ }\textbf {\bibinfo {volume} {304}},\ \bibinfo {pages}
  {1478} (\bibinfo {year} {2004})}\BibitemShut {NoStop}%
\bibitem [{\citenamefont {Mikami}\ \emph {et~al.}(2005)\citenamefont {Mikami},
  \citenamefont {Li}, \citenamefont {Fukuoka},\ and\ \citenamefont
  {Kobayashi}}]{mikami-prl-05}%
  \BibitemOpen
  \bibfield  {author} {\bibinfo {author} {\bibfnamefont {H.}~\bibnamefont
  {Mikami}}, \bibinfo {author} {\bibfnamefont {Y.}~\bibnamefont {Li}}, \bibinfo
  {author} {\bibfnamefont {K.}~\bibnamefont {Fukuoka}}, \ and\ \bibinfo
  {author} {\bibfnamefont {T.}~\bibnamefont {Kobayashi}},\ }\href@noop {}
  {\bibfield  {journal} {\bibinfo  {journal} {Phys. Rev. Lett.}\ }\textbf
  {\bibinfo {volume} {95}},\ \bibinfo {pages} {150404} (\bibinfo {year}
  {2005})}\BibitemShut {NoStop}%
\bibitem [{\citenamefont {Resch}\ \emph {et~al.}(2005)\citenamefont {Resch},
  \citenamefont {Walther},\ and\ \citenamefont {Zeilinger}}]{resch-prl-05}%
  \BibitemOpen
  \bibfield  {author} {\bibinfo {author} {\bibfnamefont {K.~J.}\ \bibnamefont
  {Resch}}, \bibinfo {author} {\bibfnamefont {P.}~\bibnamefont {Walther}}, \
  and\ \bibinfo {author} {\bibfnamefont {A.}~\bibnamefont {Zeilinger}},\
  }\href@noop {} {\bibfield  {journal} {\bibinfo  {journal} {Phys. Rev. Lett.}\
  }\textbf {\bibinfo {volume} {94}},\ \bibinfo {pages} {070402} (\bibinfo
  {year} {2005})}\BibitemShut {NoStop}%
\bibitem [{\citenamefont {Laflamme}\ \emph {et~al.}(1998)\citenamefont
  {Laflamme}, \citenamefont {Knill}, \citenamefont {Zurek}, \citenamefont
  {Catasti},\ and\ \citenamefont {Mariappan}}]{laflamme-proc-98}%
  \BibitemOpen
  \bibfield  {author} {\bibinfo {author} {\bibfnamefont {R.}~\bibnamefont
  {Laflamme}}, \bibinfo {author} {\bibfnamefont {E.}~\bibnamefont {Knill}},
  \bibinfo {author} {\bibfnamefont {W.~H.}\ \bibnamefont {Zurek}}, \bibinfo
  {author} {\bibfnamefont {P.}~\bibnamefont {Catasti}}, \ and\ \bibinfo
  {author} {\bibfnamefont {S.~V.~S.}\ \bibnamefont {Mariappan}},\ }\href@noop
  {} {\bibfield  {journal} {\bibinfo  {journal} {Proc. Roy. Soc. A}\ }\textbf
  {\bibinfo {volume} {356}},\ \bibinfo {pages} {1941} (\bibinfo {year}
  {1998})}\BibitemShut {NoStop}%
\bibitem [{\citenamefont {Nelson}\ \emph {et~al.}(2000)\citenamefont {Nelson},
  \citenamefont {Cory},\ and\ \citenamefont {Lloyd}}]{nelson-pra-00}%
  \BibitemOpen
  \bibfield  {author} {\bibinfo {author} {\bibfnamefont {R.~J.}\ \bibnamefont
  {Nelson}}, \bibinfo {author} {\bibfnamefont {D.~G.}\ \bibnamefont {Cory}}, \
  and\ \bibinfo {author} {\bibfnamefont {S.}~\bibnamefont {Lloyd}},\
  }\href@noop {} {\bibfield  {journal} {\bibinfo  {journal} {Phys. Rev. A}\
  }\textbf {\bibinfo {volume} {61}},\ \bibinfo {pages} {022106} (\bibinfo
  {year} {2000})}\BibitemShut {NoStop}%
\bibitem [{\citenamefont {Teklemariam}\ \emph {et~al.}(2002)\citenamefont
  {Teklemariam}, \citenamefont {Fortunato}, \citenamefont {Pravia},
  \citenamefont {Sharf}, \citenamefont {Havel}, \citenamefont {Cory},
  \citenamefont {Bhattaharyya},\ and\ \citenamefont
  {Hou}}]{teklemariam-pra-02}%
  \BibitemOpen
  \bibfield  {author} {\bibinfo {author} {\bibfnamefont {G.}~\bibnamefont
  {Teklemariam}}, \bibinfo {author} {\bibfnamefont {E.~M.}\ \bibnamefont
  {Fortunato}}, \bibinfo {author} {\bibfnamefont {M.~A.}\ \bibnamefont
  {Pravia}}, \bibinfo {author} {\bibfnamefont {Y.}~\bibnamefont {Sharf}},
  \bibinfo {author} {\bibfnamefont {T.~F.}\ \bibnamefont {Havel}}, \bibinfo
  {author} {\bibfnamefont {D.~G.}\ \bibnamefont {Cory}}, \bibinfo {author}
  {\bibfnamefont {A.}~\bibnamefont {Bhattaharyya}}, \ and\ \bibinfo {author}
  {\bibfnamefont {J.}~\bibnamefont {Hou}},\ }\href@noop {} {\bibfield
  {journal} {\bibinfo  {journal} {Phys. Rev. A}\ }\textbf {\bibinfo {volume}
  {66}},\ \bibinfo {pages} {012309} (\bibinfo {year} {2002})}\BibitemShut
  {NoStop}%
\bibitem [{\citenamefont {Kawamura}\ \emph {et~al.}(2006)\citenamefont
  {Kawamura}, \citenamefont {Morimoto}, \citenamefont {Mori}, \citenamefont
  {Sawae}, \citenamefont {Takarabe},\ and\ \citenamefont
  {Manmoto}}]{kawamura-ijqc-06}%
  \BibitemOpen
  \bibfield  {author} {\bibinfo {author} {\bibfnamefont {M.}~\bibnamefont
  {Kawamura}}, \bibinfo {author} {\bibfnamefont {T.}~\bibnamefont {Morimoto}},
  \bibinfo {author} {\bibfnamefont {Y.}~\bibnamefont {Mori}}, \bibinfo {author}
  {\bibfnamefont {R.}~\bibnamefont {Sawae}}, \bibinfo {author} {\bibfnamefont
  {K.}~\bibnamefont {Takarabe}}, \ and\ \bibinfo {author} {\bibfnamefont
  {Y.}~\bibnamefont {Manmoto}},\ }\href@noop {} {\bibfield  {journal} {\bibinfo
   {journal} {Int. J. Qtm. Chem.}\ }\textbf {\bibinfo {volume} {106}},\
  \bibinfo {pages} {3108} (\bibinfo {year} {2006})}\BibitemShut {NoStop}%
\bibitem [{\citenamefont {Peng}\ \emph {et~al.}(2010)\citenamefont {Peng},
  \citenamefont {Zhang}, \citenamefont {Du},\ and\ \citenamefont
  {Suter}}]{peng-pra-10}%
  \BibitemOpen
  \bibfield  {author} {\bibinfo {author} {\bibfnamefont {X.}~\bibnamefont
  {Peng}}, \bibinfo {author} {\bibfnamefont {J.}~\bibnamefont {Zhang}},
  \bibinfo {author} {\bibfnamefont {J.}~\bibnamefont {Du}}, \ and\ \bibinfo
  {author} {\bibfnamefont {D.}~\bibnamefont {Suter}},\ }\href@noop {}
  {\bibfield  {journal} {\bibinfo  {journal} {Phys. Rev. A}\ }\textbf {\bibinfo
  {volume} {81}},\ \bibinfo {pages} {042327} (\bibinfo {year}
  {2010})}\BibitemShut {NoStop}%
\bibitem [{\citenamefont {Gao}\ \emph {et~al.}(2013)\citenamefont {Gao},
  \citenamefont {Zhou}, \citenamefont {Zou}, \citenamefont {Peng},\ and\
  \citenamefont {Du}}]{gao-pra-13}%
  \BibitemOpen
  \bibfield  {author} {\bibinfo {author} {\bibfnamefont {Y.}~\bibnamefont
  {Gao}}, \bibinfo {author} {\bibfnamefont {H.}~\bibnamefont {Zhou}}, \bibinfo
  {author} {\bibfnamefont {D.}~\bibnamefont {Zou}}, \bibinfo {author}
  {\bibfnamefont {X.}~\bibnamefont {Peng}}, \ and\ \bibinfo {author}
  {\bibfnamefont {J.}~\bibnamefont {Du}},\ }\href@noop {} {\bibfield  {journal}
  {\bibinfo  {journal} {Phys. Rev. A}\ }\textbf {\bibinfo {volume} {87}},\
  \bibinfo {pages} {032335} (\bibinfo {year} {2013})}\BibitemShut {NoStop}%
\bibitem [{\citenamefont {Dogra}\ \emph {et~al.}(2015)\citenamefont {Dogra},
  \citenamefont {Dorai},\ and\ \citenamefont {Arvind}}]{dogra-pra-15}%
  \BibitemOpen
  \bibfield  {author} {\bibinfo {author} {\bibfnamefont {S.}~\bibnamefont
  {Dogra}}, \bibinfo {author} {\bibfnamefont {K.}~\bibnamefont {Dorai}}, \ and\
  \bibinfo {author} {\bibnamefont {Arvind}},\ }\href@noop {} {\bibfield
  {journal} {\bibinfo  {journal} {Phys. Rev. A}\ }\textbf {\bibinfo {volume}
  {91}},\ \bibinfo {pages} {022312} (\bibinfo {year} {2015})}\BibitemShut
  {NoStop}%
\bibitem [{\citenamefont {Devi}\ \emph {et~al.}(2012)\citenamefont {Devi},
  \citenamefont {Sudha},\ and\ \citenamefont {Rajagopal}}]{ushadevi-qip-12}%
  \BibitemOpen
  \bibfield  {author} {\bibinfo {author} {\bibfnamefont {A.~R.~U.}\
  \bibnamefont {Devi}}, \bibinfo {author} {\bibnamefont {Sudha}}, \ and\
  \bibinfo {author} {\bibfnamefont {A.~K.}\ \bibnamefont {Rajagopal}},\
  }\href@noop {} {\bibfield  {journal} {\bibinfo  {journal} {Quant. Inf.
  Proc.}\ }\textbf {\bibinfo {volume} {11}},\ \bibinfo {pages} {685} (\bibinfo
  {year} {2012})}\BibitemShut {NoStop}%
\bibitem [{\citenamefont {Sudha}\ \emph {et~al.}(2012)\citenamefont {Sudha},
  \citenamefont {Devi},\ and\ \citenamefont {Rajagopal}}]{sudha-pra-12}%
  \BibitemOpen
  \bibfield  {author} {\bibinfo {author} {\bibnamefont {Sudha}}, \bibinfo
  {author} {\bibfnamefont {A.~R.~U.}\ \bibnamefont {Devi}}, \ and\ \bibinfo
  {author} {\bibfnamefont {A.~K.}\ \bibnamefont {Rajagopal}},\ }\href@noop {}
  {\bibfield  {journal} {\bibinfo  {journal} {Phys. Rev. A}\ }\textbf {\bibinfo
  {volume} {85}},\ \bibinfo {pages} {012103} (\bibinfo {year}
  {2012})}\BibitemShut {NoStop}%
\bibitem [{\citenamefont {Das}\ \emph {et~al.}(2015)\citenamefont {Das},
  \citenamefont {Sengupta},\ and\ \citenamefont {Arvind}}]{das-qph-15}%
  \BibitemOpen
  \bibfield  {author} {\bibinfo {author} {\bibfnamefont {D.}~\bibnamefont
  {Das}}, \bibinfo {author} {\bibfnamefont {R.}~\bibnamefont {Sengupta}}, \
  and\ \bibinfo {author} {\bibnamefont {Arvind}},\ }\href@noop {} {\bibfield
  {journal} {\bibinfo  {journal} {ArXiv e-prints,}\ } (\bibinfo {year}
  {2015})},\ \Eprint {http://arxiv.org/abs/1504.02991} {1504.02991}
  \BibitemShut {NoStop}%
\bibitem [{\citenamefont {Gisin}(1996)}]{gisin-pla-96}%
  \BibitemOpen
  \bibfield  {author} {\bibinfo {author} {\bibfnamefont {N.}~\bibnamefont
  {Gisin}},\ }\href@noop {} {\bibfield  {journal} {\bibinfo  {journal} {Phys.
  Lett. A}\ }\textbf {\bibinfo {volume} {210}},\ \bibinfo {pages} {151}
  (\bibinfo {year} {1996})}\BibitemShut {NoStop}%
\bibitem [{\citenamefont {Verstraete}\ and\ \citenamefont
  {Wolf}(2002)}]{prl-verstraete-02}%
  \BibitemOpen
  \bibfield  {author} {\bibinfo {author} {\bibfnamefont {F.}~\bibnamefont
  {Verstraete}}\ and\ \bibinfo {author} {\bibfnamefont {M.~M.}\ \bibnamefont
  {Wolf}},\ }\href@noop {} {\bibfield  {journal} {\bibinfo  {journal} {Phys.
  Rev. Lett.}\ }\textbf {\bibinfo {volume} {89}},\ \bibinfo {pages} {170401}
  (\bibinfo {year} {2002})}\BibitemShut {NoStop}%
\bibitem [{\citenamefont {Cory}\ \emph {et~al.}(1998)\citenamefont {Cory},
  \citenamefont {Price},\ and\ \citenamefont {Havel}}]{cory-physicad}%
  \BibitemOpen
  \bibfield  {author} {\bibinfo {author} {\bibfnamefont {D.}~\bibnamefont
  {Cory}}, \bibinfo {author} {\bibfnamefont {M.}~\bibnamefont {Price}}, \ and\
  \bibinfo {author} {\bibfnamefont {T.}~\bibnamefont {Havel}},\ }\href@noop {}
  {\bibfield  {journal} {\bibinfo  {journal} {Physica D}\ }\textbf {\bibinfo
  {volume} {120}},\ \bibinfo {pages} {82} (\bibinfo {year} {1998})}\BibitemShut
  {NoStop}%
\bibitem [{\citenamefont {Chuang}\ \emph {et~al.}(1998)\citenamefont {Chuang},
  \citenamefont {Gershenfeld}, \citenamefont {Kubinec},\ and\ \citenamefont
  {Leung}}]{chuang-proc-98}%
  \BibitemOpen
  \bibfield  {author} {\bibinfo {author} {\bibfnamefont {I.~L.}\ \bibnamefont
  {Chuang}}, \bibinfo {author} {\bibfnamefont {N.}~\bibnamefont {Gershenfeld}},
  \bibinfo {author} {\bibfnamefont {M.}~\bibnamefont {Kubinec}}, \ and\
  \bibinfo {author} {\bibfnamefont {D.~W.}\ \bibnamefont {Leung}},\ }\href@noop
  {} {\bibfield  {journal} {\bibinfo  {journal} {Proc. Roy. Soc. A}\ }\textbf
  {\bibinfo {volume} {454}},\ \bibinfo {pages} {447} (\bibinfo {year}
  {1998})}\BibitemShut {NoStop}%
\bibitem [{\citenamefont {Long}\ \emph {et~al.}(2001)\citenamefont {Long},
  \citenamefont {Yan},\ and\ \citenamefont {Sun}}]{long-joptb}%
  \BibitemOpen
  \bibfield  {author} {\bibinfo {author} {\bibfnamefont {G.}~\bibnamefont
  {Long}}, \bibinfo {author} {\bibfnamefont {H.}~\bibnamefont {Yan}}, \ and\
  \bibinfo {author} {\bibfnamefont {Y.}~\bibnamefont {Sun}},\ }\href@noop {}
  {\bibfield  {journal} {\bibinfo  {journal} {J. Opt. B}\ }\textbf {\bibinfo
  {volume} {3}},\ \bibinfo {pages} {376} (\bibinfo {year} {2001})}\BibitemShut
  {NoStop}%
\bibitem [{\citenamefont {Leskowitz}\ and\ \citenamefont
  {Mueller}(2004)}]{leskowitz-pra-04}%
  \BibitemOpen
  \bibfield  {author} {\bibinfo {author} {\bibfnamefont {G.~M.}\ \bibnamefont
  {Leskowitz}}\ and\ \bibinfo {author} {\bibfnamefont {L.~J.}\ \bibnamefont
  {Mueller}},\ }\href@noop {} {\bibfield  {journal} {\bibinfo  {journal} {Phys.
  Rev. A}\ }\textbf {\bibinfo {volume} {69}},\ \bibinfo {pages} {052302}
  (\bibinfo {year} {2004})}\BibitemShut {NoStop}%
\bibitem [{\citenamefont {Uhlmann}(1976)}]{uhlmann-fidelity}%
  \BibitemOpen
  \bibfield  {author} {\bibinfo {author} {\bibfnamefont {A.}~\bibnamefont
  {Uhlmann}},\ }\href@noop {} {\bibfield  {journal} {\bibinfo  {journal} {Rep.
  Math. Phys.}\ }\textbf {\bibinfo {volume} {9}},\ \bibinfo {pages} {273}
  (\bibinfo {year} {1976})}\BibitemShut {NoStop}%
\bibitem [{\citenamefont {Jozsa}(1994)}]{jozsa-fidelity}%
  \BibitemOpen
  \bibfield  {author} {\bibinfo {author} {\bibfnamefont {R.}~\bibnamefont
  {Jozsa}},\ }\href@noop {} {\bibfield  {journal} {\bibinfo  {journal} {J. Mod.
  Opt.}\ }\textbf {\bibinfo {volume} {41}},\ \bibinfo {pages} {2315} (\bibinfo
  {year} {1994})}\BibitemShut {NoStop}%
\bibitem [{\citenamefont {Linden}\ and\ \citenamefont
  {Wootters}(2002)}]{linden-prl-2-02}%
  \BibitemOpen
  \bibfield  {author} {\bibinfo {author} {\bibfnamefont {N.}~\bibnamefont
  {Linden}}\ and\ \bibinfo {author} {\bibfnamefont {W.~K.}\ \bibnamefont
  {Wootters}},\ }\href@noop {} {\bibfield  {journal} {\bibinfo  {journal}
  {Phys. Rev. Lett.}\ }\textbf {\bibinfo {volume} {89}},\ \bibinfo {pages}
  {277906} (\bibinfo {year} {2002})}\BibitemShut {NoStop}%
\end{thebibliography}
%
\end{document}